\documentclass[prb,showpacs,preprint,amsmath,amssymb,groupadress,superscriptaddress]{revtex4}
\usepackage{graphicx}
\begin{document}

\newcommand{\HG}{HgCr$_2$S$_4$}

\title{Non-collinear long-range magnetic ordering in \HG}

\author{L.C. Chapon}
\affiliation{ISIS facility, Rutherford Appleton Laboratory-CCLRC,
Chilton, Didcot, Oxfordshire, OX11 0QX, United Kingdom. }
\author{P.G. Radaelli}
\affiliation{ISIS facility, Rutherford Appleton Laboratory-CCLRC,
Chilton, Didcot, Oxfordshire, OX11 0QX, United Kingdom. }
\affiliation{Dept. of Physics and Astronomy, University College
London, Gower Street, London WC1E 6BT, United Kingdom}
\author{Y. S. Hor}
\affiliation{Materials Science Division, Argonne National
Laboratory, Argonne, IL  60439}
\author{M.T.F. Telling}
\affiliation{ISIS facility, Rutherford Appleton Laboratory-CCLRC,
Chilton, Didcot, Oxfordshire, OX11 0QX, United Kingdom. }
\author{J.F. Mitchell}
\affiliation{Materials Science Division, Argonne National
Laboratory, Argonne, IL  60439}

\begin{abstract}
The low-temperature magnetic structure of \HG\ has been studied by
high-resolution powder neutron diffraction. Long-range
incommensurate magnetic order sets in at T$_N\sim$22K with
propagation vector \textbf{k}=(0,0,$\sim$0.18).   On cooling below
T$_N$, the propagation vector increases and saturates at the
commensurate value \textbf{k}=(0,0,0.25).  The magnetic structure
below T$_N$ consists of ferromagnetic layers in the
\textit{ab}-plane stacked in a spiral arrangement along the
\textit{c}-axis. Symmetry analysis using corepresentations theory
reveals a point group symmetry in the ordered magnetic phase of 422
(D$_4$), which is incompatible with macroscopic ferroelectricity.
This finding indicates that the spontaneous electric polarization
observed experimentally cannot be coupled to the magnetic order
parameter.
\end{abstract}


\maketitle

There has been a considerable renewal of interest in multiferroic
materials in the last few years, in particular in magnetoelectric
oxides such as TbMnO$_3$\cite{PhysRevLett.95.087206},
TbMn$_2$O$_5$\cite{PhysRevLett.96.097601,PhysRevLett.93.177402} or
the Kagom\'{e} staircase compound
Ni$_3$V$_2$O$_8$\cite{PhysRevLett.93.247201}. These systems are all
improper ferroelectrics and exhibit as a common feature strong
magnetic frustration due to the presence of competing interactions
or geometrical constraints.  Consequently, their magnetically
ordered states are typically quite complex. Independent of the
microscopic mechanism responsible for ferroelectricity-antisymmetric
Dzyaloshinskii-Moriya interaction or symmetric exchange
striction-the symmetry breaking process associated with long-range
magnetic order leaves a non-centrosymmetric structural point group,
allowing a macroscopic spontaneous polarization to develop below the
magnetic critical temperature. Another common element of fundamental
interest is the nature of magnetic ordering itself, which involves
propagation vectors inside the Brillouin zone, whether the magnetic
order is commensurate or not with respect to the crystalline unit
cell. In all these cases, it is crucial to perform
symmetry analysis using co-representation theory to determine the point-group symmetry lowering induced by magnetic ordering \cite{RadaelliSymmetry}. \\
\indent Recently, attention has been drawn to chalcogenide chromium
spinels of the type ACr$_2$X$_4$ (A=Cd,Hg; X=S,Se) which are weakly
ferroelectric in their magnetically ordered state and have been
classify as multiferroic
materials\cite{CdCr2S4Nature,PhysRevLett.96.157202,PhysRevB.73.224442}.
However, the dielectric behaviors are reminiscent of relaxor
ferroelectrics, in which polar domains of nanometric sizes appear
below the critical temperature.  Although most chalcogenide spinels
are Heisenberg ferromagnets, \HG\ shows a more complex magnetic
behavior at low temperature, indicating stronger competition between
short- and long-range interactions: despite a large positive Weiss
temperature of 142K\cite{PhysRev.151.367}, indicating dominant
ferromagnetic interactions, the low temperature magnetic structure
found by neutron diffraction experiments \cite{MagneticHgCr2S4}
corresponds to a non collinear antiferromagnet, in agreement with
recent bulk magnetization measurements\cite{PhysRevB.73.224442}.
However, Weber and co-workers noted in \cite{PhysRevLett.96.157202}
that \textit{the true nature of the magnetic state of \HG\ still
needs some clarification} due to discrepancies between recent
magnetization results \cite{PhysRevB.73.224442} and the earlier
neutron \cite{MagneticHgCr2S4} and optical studies
\cite{PhysRevB.1.319}. Specifically, these earlier reports claim a
much higher transition temperature than that of Weber et al.
Determining the exact magnetic ground state, including its symmetry
properties, is of crucial importance to study the potential coupling
of magnetic and ferroelectric order
parameters and is the prime motivation of the present work.\\
\indent In this Communication, we report a complete analysis of the
magnetic structure of \HG\ studied by high-resolution neutron powder
diffraction. Our results show that T$_N=$22.5K, in agreement with
recent magnetization and specific heat
measurements\cite{PhysRevB.73.224442} and contradicting early
neutron diffraction and optical
studies\cite{MagneticHgCr2S4,PhysRevB.1.319} that reported a
transition temperature around 60 K.  The magnetic structure is
incommensurate at T$_N$, but the propagation vector rapidly evolves
with decreasing temperature, becoming nearly commensurate at 1.5 K.
 We also report the symmetry properties of the low-temperature magnetic
structure as analyzed using corepresentation theory\cite{Kovalev}.
We find that only a single mode belonging to a two-dimensional
irreducible co-representation of the little group fits the
experimental data.  The resulting ordered magnetic structure refined
from the neutron data is a spiral with propagation vector parallel
to its axis. Symmetry analysis reveals that for \textbf{k}
incommensurate the structural point group associated with this
magnetic configuration is $422$, forbidding the development of a
macroscopic polar vector at T$_N$. For $\textbf{k} = \frac{1}{4}$
exactly, the point group symmetry is lower ($222$ or $2$, depending
on the choice of the overall phase). However, this situation is only
found at very low temperatures, leading to the clear conclusion that
the appearance of hysteresis loops in the electric spontaneous
polarization below 70K\cite{PhysRevLett.96.157202} $\gg$ T$_N$ is not related to antiferromagnetic order.\\
\indent Polycrystalline \HG\ was synthesized by combining powders of
Cr (99.99\%) and S (99.999\%) with liquid Hg (99.9998\%) mixed in a
stoichiometric ratio. The mixture was sealed in an evacuated quartz
tube and heated slowly to 600 $^{\circ}$C for several days. The
powder was then ground, pelletized, and sealed again for sintering
at 800 $^{\circ}$C for several weeks. Powder neutron diffraction
data were collected on the high-resolution OSIRIS diffractometer at
the ISIS pulsed neutron source (UK). A 1.5g sample was enclosed in a
vanadium can placed in a helium cryostat. Long scans were recorded
above (30K) and below (1.5K) T$_N$ to solve the magnetic structure.
Shorter scans in a narrower d-spacing range (5.2-6.4\AA) were
recorded between 1.5K and 25.5K in 1K steps to follow the
temperature dependence of the (111)+\textbf{k} Bragg peak. Rietveld
refinements were performed with the program FullProF\cite{Fullprof}.
Symmetry analysis using representation theory was performed with the
help of the software MODY\cite{Mody}. Co-representation matrices and
modes are obtained from the corresponding irreducible
representations as described by Kovalev \cite{Kovalev}.
Results are presented using Kovalev's notation. We label representations as $\Gamma_i$ and corepresentations as $\widetilde{\Gamma}_i$.\\

\begin{table}[ph!]
\begin{scriptsize}
\begin{center}
\begin{tabular}[b]{c|*{4}{c}}
\hline \hline
 Atoms & Cr$_1$ & Cr$_2$ & Cr$_3$ & Cr$_4$  \\
 \hline
$\widetilde{\Gamma}_5(1-1)$ & $\epsilon^2$*(1,i,0) & (1,i,0) &
$\epsilon$*(1,i,0) & $\epsilon^3$*(1,i,0) \\
$\widetilde{\Gamma}_5(1-2)$ & $\epsilon^2$*(1,-i,0) & (1,-i,0) &
$\epsilon$*(1,-i,0) & $\epsilon^3$*(1,-i,0) \\
$\widetilde{\Gamma}_5(2-1)$ & $\epsilon^2$*(-i,1,0) & (-i,1,0) &
$\epsilon$*(i,-1,0) & $\epsilon^3$*(i,-1,0) \\
$\widetilde{\Gamma}_5(2-2)$ & $\epsilon^2$*(i,1,0) & (i,1,0) &
$\epsilon$*(-i,-1,0) & $\epsilon^3$*(-i,-1,0) \\
\end{tabular}
\end{center}
\end{scriptsize}
\label{irrep1} \caption{Symmetrized in-plane basis vectors for the
co-representations $\widetilde{\Gamma}_5$. $\epsilon$
=e$^{-i\frac{\pi}{2}\delta}$, and $\epsilon$* is the complex
conjugate of $\epsilon$. The atomic positions of Cr$_i$ (i=1,4) are
given in the text.}
\end{table}

The neutron powder diffraction pattern collected at 30 K is
consistent with the cubic crystal structure
\cite{HgCr2S4crystalstructure}, space group
\textit{Fd$\overline{3}$m}\footnote{The structure is described in
the second setting of the International Tables of Crystallography}.
The high neutron absorption cross-section of Hg and reduced Q-range
intrinsic to this instrument prevented us from extracting reliable
thermal parameters and from probing possible off-center displacement
of the Cr atoms as recently suggested \cite{PhysRevB.73.224442}. A
detailed temperature-dependent structural investigation by X-ray
diffraction is currently underway and will be reported elsewhere. At
1.5K, 12 magnetic Bragg peaks appear at low-Q in the diffraction
pattern; all can
 be indexed with the propagation vector
\textbf{k}=(0,0,$\delta$) (with respect to the conventional
F-centered cell) in agreement with a previous
study\cite{MagneticHgCr2S4}. $\delta \sim \frac{1}{4}$ within our
experimental resolution, i.e. the periodicity of the magnetic
structure is commensurate with
respect to the nuclear unit-cell at 1.5 K. \\
\indent The symmetry-allowed magnetic
structures have been determined by co-representation analysis. Here
we consider four Cr atoms in the primitive unit-cell (labelled Cr$_i$
i=1,4) respectively positioned at
($\frac{1}{2}$,$\frac{1}{2}$,$\frac{1}{2}$),
($\frac{1}{4}$,$\frac{3}{4}$,0), ($\frac{3}{4}$,0,$\frac{1}{4}$),
(0,$\frac{1}{4}$,$\frac{3}{4}$).  The remaining 12 atoms are generated
by  F-centering operations. The magnetic
representation $\Gamma$ of dimension 12 is reduced into the direct
sum of irreducible representations:
\begin{equation}
\Gamma=\Gamma_1^1+2\Gamma_2^1+2\Gamma_3^1+\Gamma_4^1+3\Gamma_5^2
\end{equation}
where the subscripts label the irreducible representations following
Kovalev's notation\cite{Kovalev} and the superscripts refer to their
dimensionality. The magnetic transition is continuous, and according
to the Landau theory of phase transitions\cite{Landau}, should
involve a single irreducible representation. Indeed, the magnetic
structure is found to be consistent only with $\Gamma_5$, with modes
in the \textit{ab}-plane. The symmetrized in-plane basis vectors for
the associated co-representation $\widetilde{\Gamma}_5$ are reported
in Table I. Only the modes labelled $\widetilde{\Gamma}_5(1-1)$ and
$\widetilde{\Gamma}_5(1-2)$ fit the data. Both modes correspond to
spiral arrangements that only differ by their rotation direction, as
seen from the sign of the imaginary part of the Fourier
coefficients. The phases for atoms 1-4 are fixed by symmetry and are
directly related to the fractional position along \textit{c} in such
a way that atoms belonging to a same plane (including atoms
generated by F-centering translations) have the same phase. The
magnetic structure is therefore modeled by a single parameter: the
magnitude of the Fourier component, i.e. the magnetic moment.
\begin{figure}[h!]
\includegraphics[scale=0.8]{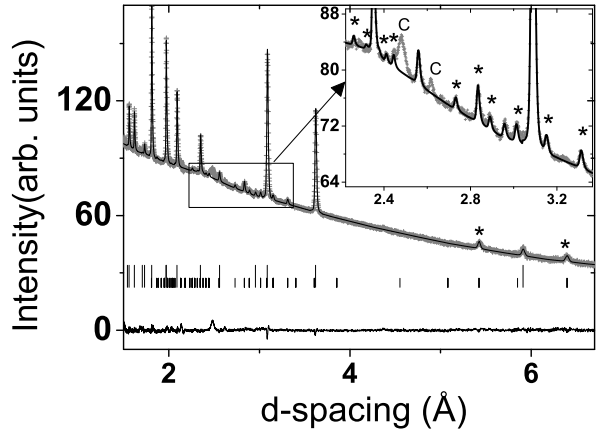}
\caption{Rietveld refinement of the neutron diffraction pattern for
\HG\ at 1.5K. The grey crosses represent the experimental data while
the black solid line represents the refined model. The solid line at
the bottom is the difference curve between data and refinement. The
long (short) tick-marks show the positions of nuclear (magnetic)
Bragg reflections. Non-extinct magnetic reflections are marked by an
asterisk symbol. Two un-indexed diffraction peak, marked by "C",
correspond to parasitic scattering from the cryostat.}
\end{figure}

\indent The refinement of the neutron diffraction pattern at 1.5K is
presented in Fig. 1, showing essentially perfect agreement between
experimental data and model. As mentioned before, the magnetic
structure shown in Fig. 2 corresponds to a simple spiral propagating
along \textit{c} and ferromagnetic arrangements in the
\textit{ab}-plane. Within a tetrahedral Cr$_4$ unit of the
pyrochlore lattice, the angle between first-neighbor spins along
\textit{c} is $\alpha=\pi\delta/2$, i.e. 22.5 $^{\circ}$ at 1.5K.
The magnetic moment of 2.7(1) $\mu_B$ is slightly smaller than the
expected spin-only contribution for Cr$^{3+}$, S=3/2. This value
agrees well with previous studies \cite{MagneticHgCr2S4} and seems
to indicate that a small fraction of the magnetic moment remain
disordered at low temperature. Nevertheless, one must take this
result with care due to the strong correlation between magnitude of
the moment and
absorption correction in the refinement.\\
\indent On warming, the propagation vector deviates from
$\delta=\frac{1}{4}$ at 4K, which corresponds to a monotonous
variation of the angle $\alpha$, and its magnitude decreases rapidly
at higher temperature (Fig. 3). It is difficult to ascribe the low
temperature result where $\delta=\frac{1}{4}$ to a lock-in
transition, due to the narrow temperature region in which this
behavior occurs. Nonetheless, there is a clear tendency towards
saturation at this value. The variation of the intensity of the
(111)+\textbf{k} magnetic reflection, presented in Fig. 3.,
indicates a N\'{e}el transition around 22.5K, in excellent agreement
with recent specific heat results showing a lambda anomaly at
22K\cite{PhysRevB.73.224442} and the strong decrease of the
magnetization at low magnetic field. However, this finding
contradicts earlier neutron and optical studies
\cite{MagneticHgCr2S4,PhysRevB.1.319}suggesting the long range
ordered magnetic state persists up to 60K. Also in contradiction
with \cite{MagneticHgCr2S4}, the magnetic order parameter obeys a
power law close to T$_N$.\\
\indent When considering the origin of the magnetic structure of
\HG\ one can eliminate direct exchange at first approximation
because of the much larger interatomic Cr-Cr distance with respect
with analogous oxide spinels. The very nature of the magnetic
structure therefore indicates that antiferromagnetic next-nearest
neighbor interactions compete directly with ferromagnetic
super-exchange Cr-S-Cr interactions. The parameter that seems to
control the nature of the magnetic ground state is the Cr-S-Cr
angle. In \HG\, this angle is 98$^{\circ}$ at room temperature
according to structural parameters derived by
\cite{PhysRevB.73.224442}, larger than for ferromagnetic
CdCr$_2$S$_4$, for which the value is 96.9$^{\circ}$
\cite{HgCr2S4crystalstructure}. This is in agreement with
superexchange theory since larger deviation from 90$^{\circ}$
angles, for which antiferromagnetic correlation due to electron
transfer is forbidden by symmetry, will increase the orbital
overlap. The apparent extreme sensitivity of magnetic structure to
angle in these compounds calls for a systematic study of the
magnetic properties as a function of the averaged Cr-S-Cr
bond angle.\\
\begin{figure}[h!]
\includegraphics[scale=0.4,angle=270]{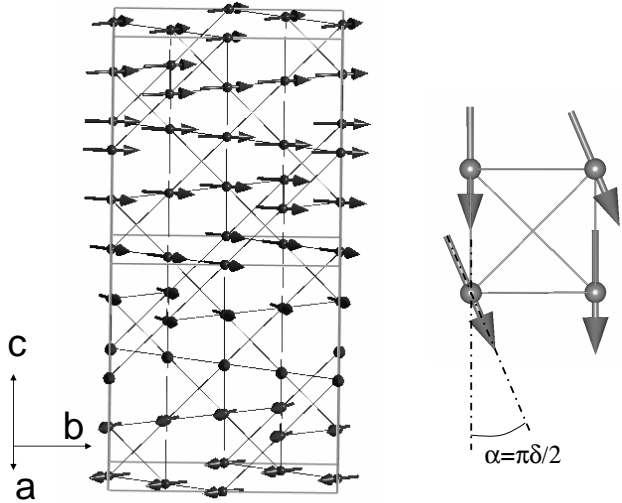}
\caption{Left: Magnetic structure at 1.5K showing a spiral
arrangement of ferromagnetic layers in the \textit{ab}-plane. Only
Cr atoms/spins are shown as grey spheres/arrows. Cr atoms are
connected by thin grey lines showing the underlying pyrochlore
lattice. Two unit-cells along the \textit{c}-direction are
represented. Right: Spin-configuration within an isolated
tetrahedral Cr$_4$ unit.}
\end{figure}
\indent Next, we consider the ferroelectricity of \HG\
\cite{PhysRevLett.96.157202} and demonstrate that it cannot arise
from coupling to the magnetic order parameter. The onset of
ferroelectricity induced by symmetry breaking associated with the
long-range magnetic order can be directly determined by symmetry
analysis of the magnetic structure. In cases where \textbf{k} lies
inside the Brillouin zone (\textbf{k} and \textbf{-k} are not
related by a reciprocal lattice vector), only the use of
co-representation theory can capture the full symmetry properties.
In Table II, we report the matrix representatives of the symmetry
operations of the little group G$_\textbf{k}$ (including
anti-unitary operations) for the irreducible corepresentations
$\widetilde{\Gamma}_5$. A symmetry operator is preserved in the
magnetically ordered phase if and only if it is equivalent to a
lattice translation (including F-centering operations). Matrices for
the proper (rotation) operations, or proper operations combined with
the complex conjugation operator K (anti-unitary rotations), are
diagonal are therefore equivalent to a translation for k
incommensurate.
\begin{figure}[h!]
\includegraphics[scale=0.8]{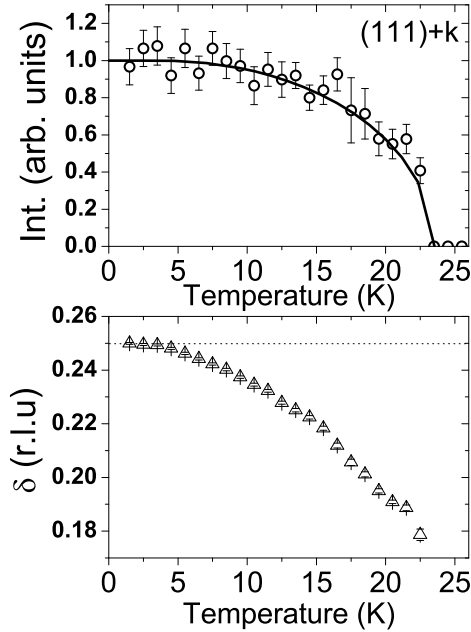}
\caption{Upper panel: Integrated intensity of the (111)+k magnetic
Bragg peak versus temperature. The solid line is a guide to the
eyes. Lower panel: temperature dependence of the propagation vector
component $\delta$ as a function of temperature. The dotted line
indicates the commensurate value $\delta$=$\frac{1}{4}$}
\end{figure}
In other words, these operations transform a basis function of the
two-dimensional subspace into itself, modulo a phase factor (which
is equivalent to a translation for incommensurate k-vectors). On the
contrary, the matrices for improper operations are off-diagonal,
which means that these operators are not preserved in the
magnetically ordered state. \indent From the list of symmetry
operators preserved, one can deduce the structural point group $422$
($D4$ in Schoenflies notation). Notably, this point group is
\textit{inconsistent} with the development of a macroscopic polar
vector. In general, the point-group symmetry is lower for exactly
commensurate \textbf{k} because not all phase factors are equivalent
to a lattice translation. Furthermore, anti-unitary operators may be
preserved or violated depending on the \emph{overall} phase factor,
which cannot be determined by neutron diffraction. Here, for
$\delta=\frac{1}{4}$, the point group is 2 for a generic phase and
222 for special choices of the overall phase. If the global phase is
zero modulo $\frac{\pi}{8}$, the two-fold axes along the [110] and
[$\overline{1}$10] directions are preserved, whereas for a phase of
$\frac{\pi}{16}$ modulo $\frac{\pi}{8}$, the two-fold axes along
[100] and [010] are preserved. The two-fold axis along the [001]
direction is always preserved.\\
\indent The condition for
ferroelectric order in non-collinear magnets has also been recently
discussed phenomenologically by M.
Mostovoy\cite{PhysRevLett.96.067601}. Mostovoy has shown that
magnetically induced ferroelectric order can develop in
non-collinear magnets only if the vector product between the spin
rotation axis \textbf{e} of the magnetic structure and propagation
vector \textbf{k} is nonzero, a condition obviously not fulfilled
here since \textbf{k} and \textbf{e} are
collinear.\\
\indent The origin of ferroelectricity and associated phenomenal
magnetoelectric properties in \HG\ remains an open question. We have
demonstrated conclusively that, symmetry breaking of the magnetic
structure does not support a ferroelectric state. However, it has
been shown that hysteresis loops of the electric polarization appear
below $\sim$70K \cite{PhysRevLett.96.157202}. It is possible that
the ferroelectric state could be coupled to short-range
ferromagnetic correlations developing around the same temperature.
The presence of nano-polar regions is in agreement with this
picture. However in this scenario, one might expect an abrupt
decrease of the spontaneous polarization below the magnetic ordering
transition, due to the strong reduction of ferromagnetic
fluctuations. This is apparently inconsistent with experimental
observation, since Weber et al.\cite{PhysRevLett.96.157202} report
that hysteresis loops become \textit{more pronounced with decreasing
temperature}. In the light of this temperature variation, it would
be particularly interesting to study a direct signature of
short-range ferromagnetic order with small angle neutron scattering
to explain why ferroelectric order persists below T$_N$.
Additionally, a detailed temperature dependence of the spontaneous
polarization is needed to explore the exact nature of ferroelectric
ordering.\\
\indent In summary, we find that non-collinear magnetic ordering
develops at 22.5K in \HG\ with a spiral structure. Analysis with
corepresentation theory shows that magnetic symmetry breaking is
inconsistent with the development of a macroscopic polar vector,
suggesting another mechanism-such as coupling to short-range
ferromagnetic order-as the link between magnetic and electric
properties in this material. \\
\begin{table*}[h!]
\begin{ruledtabular}
\begin{tabular}[b]{c|*{8}{c}}
 Coreps & h$_1$/Kh$_{25}$ & h$_4$/Kh$_{28}$ & h$_{14}$/Kh$_{38}$ & h$_{15}$/Kh$_{39}$ & h$_{26}$/Kh$_{2}$ & h$_{27}$/Kh$_{3}$ & h$_{37}$/Kh$_{13}$ & h$_{40}$/Kh$_{16}$ \\
 \hline
 $\widetilde{\Gamma}_5$ &
 $ \left( \begin{array}{cc}  1     &  0      \\   0     &  1      \end{array}  \right) $ &
 $ \left( \begin{array}{cc}  -1     &  0      \\   0     &  -1      \end{array}  \right) $ &
 $ \left( \begin{array}{cc}   i\epsilon     &  0      \\   0     &   -i\epsilon     \end{array}  \right) $ &
 $ \left( \begin{array}{cc}  -i \epsilon     &  0      \\   0     &  i \epsilon      \end{array}  \right) $ &
 $ \left( \begin{array}{cc}  0     &  \epsilon      \\   \epsilon     &  0      \end{array}  \right) $ &
 $ \left( \begin{array}{cc}  0     &  -\epsilon      \\   -\epsilon     &  0      \end{array}  \right) $ &
 $ \left( \begin{array}{cc}  0     &  -i      \\   i     &  0      \end{array}  \right) $ &
 $ \left( \begin{array}{cc}  0     &   i      \\  -i     &  0      \end{array}  \right) $ \\
  &
 $ \left( \begin{array}{cc}  0     &  1      \\   1     &  0      \end{array}  \right) $ &
 $ \left( \begin{array}{cc}  0     &  -1      \\   -1     &  0      \end{array}  \right) $ &
 $ \left( \begin{array}{cc}  0     &  i\epsilon*      \\   -i\epsilon*     &   0     \end{array}  \right) $ &
 $ \left( \begin{array}{cc}  0     &  -i\epsilon*      \\   i\epsilon*     &   0      \end{array}  \right) $ &
 $ \left( \begin{array}{cc}   \epsilon*     &  0      \\   0     &  \epsilon*      \end{array}  \right) $ &
 $ \left( \begin{array}{cc}  -\epsilon*     &  0      \\   0     &  -\epsilon*      \end{array}  \right) $ &
 $ \left( \begin{array}{cc}  -i     &  0      \\   0     &  i      \end{array}  \right) $ &
 $ \left( \begin{array}{cc}  i     &   0      \\  0     &  -i      \end{array}  \right) $ \\
\end{tabular}
\end{ruledtabular} \label{irrep1} \caption{Matrix representatives of
the irreducible co-representation $\widetilde{\Gamma}_5$ of the
little group G$_k$ (space group $G=Fd\overline{3}m$ and
k=$\textbf{k$_6$}$=(0 \ 0 \ $\delta$)). The labels for symmetry
operations and numbering of the irreducible representations follow
Kovalev's notation. Matrices for unitary (h) and anti-unitary (Kh)
symmetry operations are given on the first and second row,
respectively. $\epsilon$ =e$^{-i\frac{\pi}{2}\delta}$}
\end{table*}

\indent We would like to acknowledge I. Sergienko, for pointing out
an error in the first draft and for helpful discussions. Research
carried out in the Materials Science Division at Argonne National
Laboratory is funded by the US Department of Energy Office of
Science, Basic Energy Sciences, under Contract W-31-109-ENG-38.


\end{document}